\documentclass{pasj00}
\begin{document}
\SetRunningHead{T. Kato et al.}{V463 Sct: Nova with a Prominent Premaximum Halt}

\Received{}%{yyyy/mm/dd}
\Accepted{}%{yyyy/mm/dd}

\title{V463 Scuti (Nova Sct 2000): Rapidly Evolving Nova with \\ a Prominent Premaximum Halt}

\author{Taichi \textsc{Kato}, Makoto \textsc{Uemura}}
\affil{Department of Astronomy, Kyoto University,
       Sakyo-ku, Kyoto 606-8502}
\email{tkato@kusastro.kyoto-u.ac.jp, uemura@kusastro.kyoto-u.ac.jp}

\author{Katsumi \textsc{Haseda}}
\affil{Variable Star Observers League in Japan (VSOLJ), 2-7-10 Fujimidai,
       Toyohashi City, Aichi 441-8135}
\email{khaseda@mx1.tees.ne.jp}

\author{Hitoshi \textsc{Yamaoka}}
\affil{Faculty of Science, Kyushu University, Fukuoka 810-8560}
\email{yamaoka@rc.kyushu-u.ac.jp}

\author{Kesao \textsc{Takamizawa}}
\affil{Variable Star Observers League in Japan (VSOLJ), 65-1 Oohinata,
       Saku-machi, Nagano 384-0502}
\email{k-takamizawa@nifty.ne.jp}

\author{Mitsugu \textsc{Fujii}}
\affil{Fujii-Bisei Observatory, 4500 Kurosaki, Tamashima, Okayama
       713-8126}
\email{aikow@po.harenet.ne.jp}

\email{\rm{and}}

\author{Seiichiro \textsc{Kiyota}}
\affil{Variable Star Observers League in Japan (VSOLJ),
       1-401-810 Azuma, Tsukuba, 305-0031}
\email{skiyota@nias.affrc.go.jp}

%%% end:list of authors


\maketitle

\begin{abstract}
   We summarize photometric and spectroscopic observations of V463 Sct
(Nova Sct 2000), which was originally thought to be a red variable.
The spectrum taken on 2000 March 16.81 UT showed prominent emission lines
with a FHWM of 990 km s$^{-1}$ (H$\alpha$).  The light curve shows
a conspicuous premaximum halt lasting at least for 24 d, and a late-phase
flare-like maximum.  The nova then started rapidly fading at a rate
corresponding to $t_2$ = 15 $\pm$ 3 d.  Long premaximum halts have been
considered as a unique character of the ``slowest" novae.  The present
observation, however, suggests that the long premaximum halts are not
a unique character of the slowest novae, but a more general phenomenon
spreading over a wider range of nova speed classes than has been previously
believed.  A recent interpretation of premaximum halts requires that
the conditions of thermonuclear runaway was only marginally satisfied.
Since such conditions are more difficult to meet in rapidly evolving novae,
V463 Sct would provide an unique opportunity in testing this interpretation.
The early post-outburst spectrum showed co-existence of Fe\textsc{II}
lines and some forbidden lines, which suggests that substantial amount
of material may have been ejected before the observed optical maximum.
The impact of the modern global network (VSNET) on confirmatory processes
of transient objects is briefly discussed.
\end{abstract}

\section{Introduction}

   Classical novae outbursts are thermonuclear runaways
(TNR, \cite{sta87novareview}, \cite{sta99novareview}, \cite{sta00novareview})
on a mass-accreting white dwarf in cataclysmic variables (CVs)
[for a general review of CVs, see \cite{war95book}].
Observational distinction of classical novae is usually done
spectroscopically.  Classical novae can be easily distinguished from
other high-amplitude variable stars by the presence of broad emission
lines often accompanied by a P Cyg-type profile.

   Such a distinction, however, is sometimes difficult especially when
the detection of a new object on a photograph or a CCD image was done
at later times.  Difficult conditions are also easily met when discovery
alert is not timely distributed.  In some cases, researchers are obliged
to distinguish novae from other objects only with limited information.
For example, only 50\% of the reported novae in \citet{due87novaatlas}
(up to 1986) were spectroscopically confirmed.  Some objects later
turned out to be Mira-type stars (e.g. \cite{zha94lqsgr}) or large-amplitude
dwarf novae [see \citet{kat01hvvir} for a review of such objects].
Early spectroscopic confirmation based on a secure perspective of a newly
discovered object is therefore a necessary condition for modern nova
studies.  The condition has been greatly improved by the advent of
the global network of observers, VSNET,\footnote{
  $\langle$ http://www.kusastro.kyoto-u.ac.jp/vsnet/. $\rangle$
}, which enabled quick dissemination of
discovery alerts.  The best examples include V1548 Aql = Nova Aql 2001
\citep{kat01v1548aql} and V463 Sct = Nova Sct 2000 (this paper), which
may have been overlooked with a traditional confirmatory procedure.

\section{Discovery}\label{sec:discovery}

   In 2000 March, one of the authors (KH) discovered a new variable star
named Had~V46.  Because of the close presence of an IRAS source
(IRAS 18311$-$1447), this variable star was initially regarded as
a Mira-type long-period variable.  However, we noticed a small offset
of the reported position from the IRAS position.  We thereby took CCD
images in order to check the identification with the IRAS source.
The CCD images taken on 2000 March 14.9 UT showed that the IRAS source is
more likely to be identified with a red USNO star at
\timeform{18h 34m 02.25s}, \timeform{-14D 45' 33.6''} (J2000.0)
(star A in Figure \ref{fig:id})
and an unexpected fading of Had~V46.  Since Had~V46 looked like to be a
large-amplitude (probably eruptive) non-red variable star, the object was
reported as a possible nova \citep{has00v463sct}.  We examined past
photographic survey images taken by KT and KH.  The object was not
recorded down to 14.0 mag on 21 photographs between 1996 May 25 and 1999
November 10.
Furthermore, spectroscopic observation by MF revealed the presence of
strong broad emission lines, which indicates that Had~V46 is indeed
a nova \citep{uem00v463sct}.  The object was given the permanent variable
star designation of V463 Sct.

\section{Astrometry and Identification}

\begin{figure*}
  \begin{center}
%    \FigureFile(85mm,85mm){kiyota.eps} \hskip 3mm
    \FigureFile(85mm,85mm){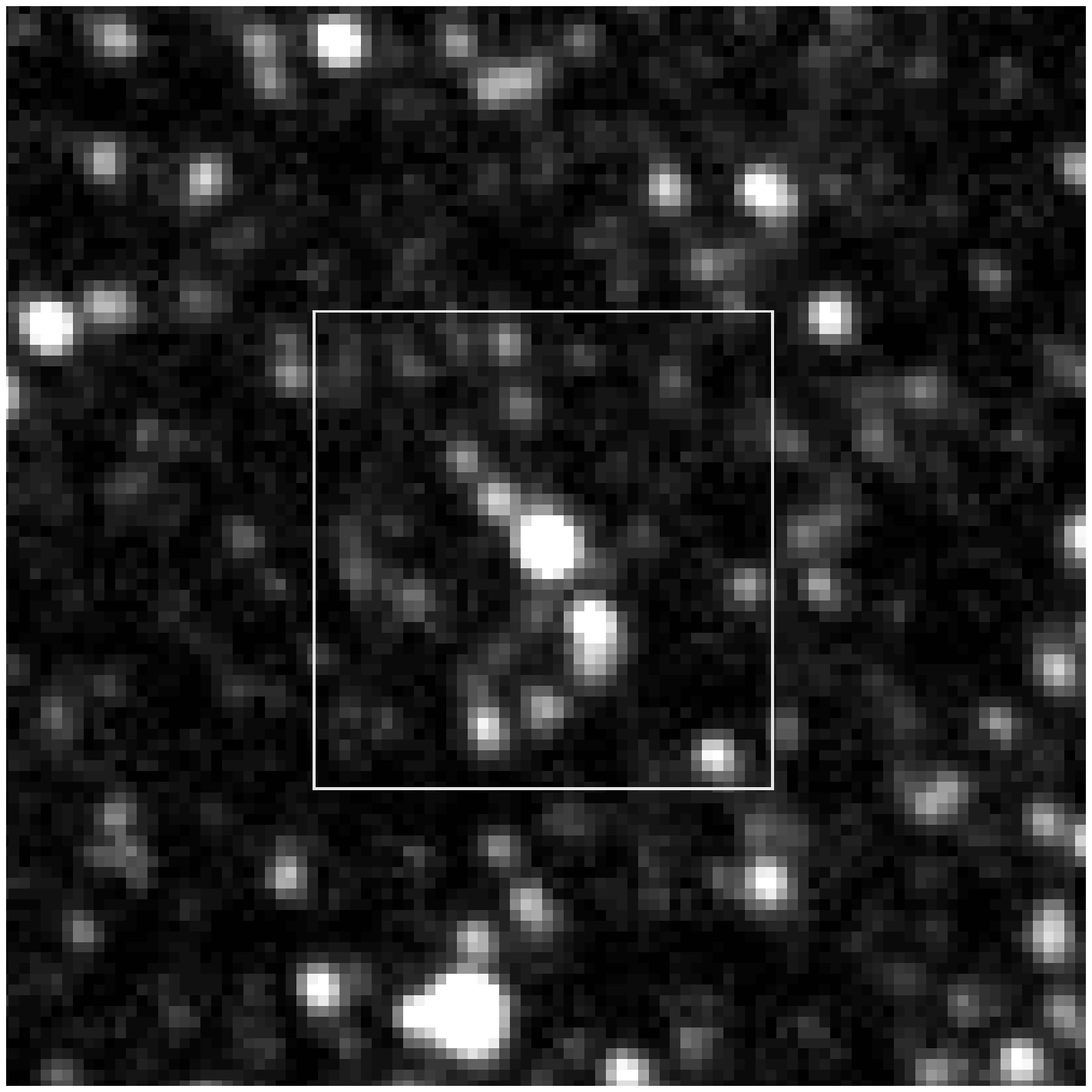} \hskip 3mm
%    \FigureFile(85mm,85mm){dss.eps}
    \FigureFile(85mm,85mm){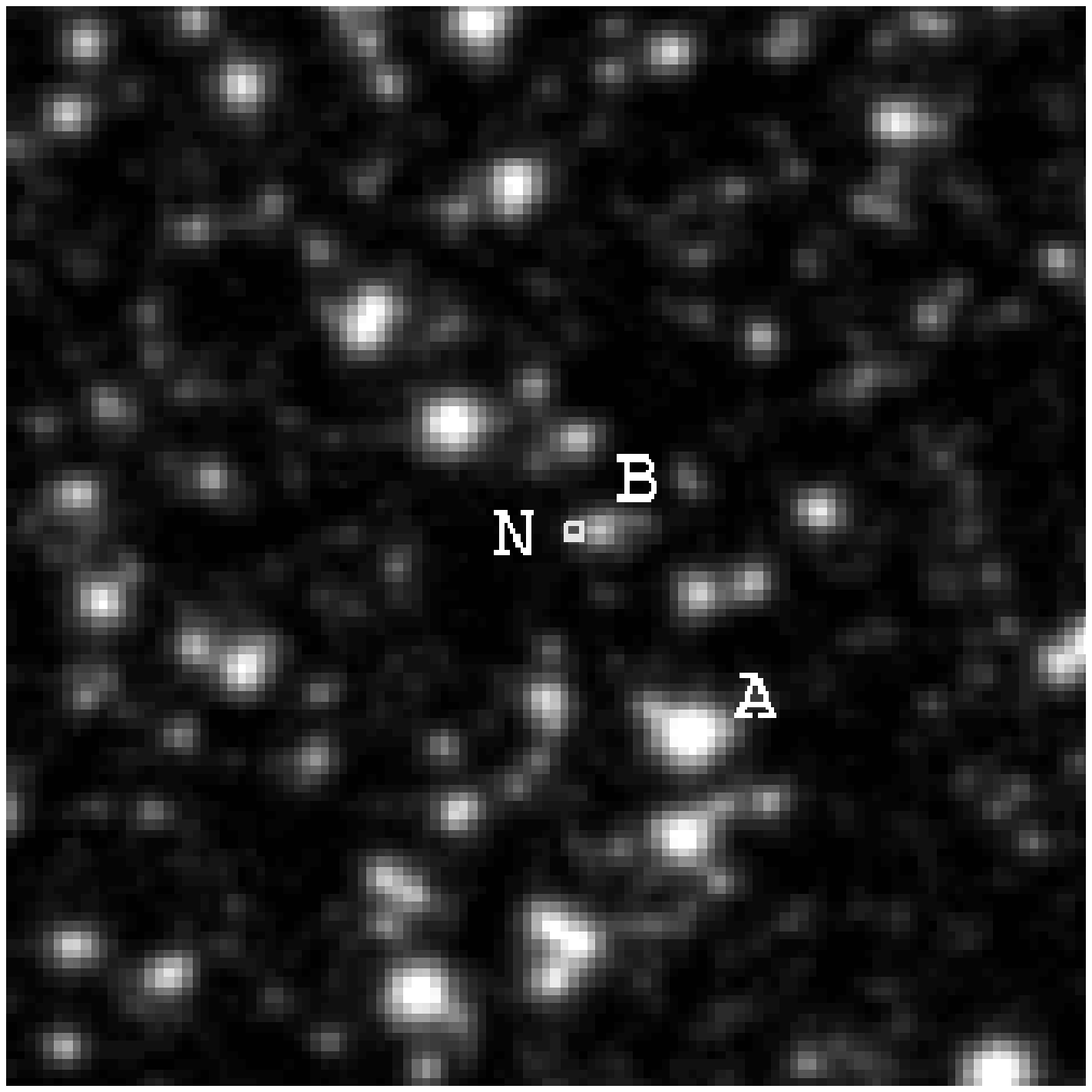}
  \end{center}
  \caption{SK's outburst image (left, 5 arcminutes square) taken on 2000
  March 16.8 UT and
  the DSS2 quiescent identification (right, 2 arcminutes square) of V463
  Sct.  North is up and east is left on both images.  The nearest USNO
  star (B in the right image) is about 3'' apart from the nova position.}
  \label{fig:id}
\end{figure*}

   Astrometry of Had~V46 was done with Kyoto images taken on 2000 March 14.9
UT and SK's image taken on March 16.8 UT, with GSC-ACT reference
stars.  The accurate position of Had~V46 was found to be 
\timeform{18h 34m 03.13s}, \timeform{-14D 45' 12.4''}
(J2000.0, internal error \timeform{0''.2})\footnote{This value supersedes
the one reported in \citet{uem00v463sct}.}.  Figure \ref{fig:id} shows a
comparison of the outburst image and quiescent identification on DSS.
The object was not detected on all available DSS images between 1984 August
and 1996 September (6 images).
The lack of detection of the prenova on these DSS images leads to an upper
limit $V \sim$ 20 for the quiescent magnitude of
V463 Sct \citep{uem00v463sct}.

\section{Photometry}

   One of the authors (SK) obtained multicolor photometry with
a 25-cm SCT telescope and an AP-7 CCD.  The magnitudes were determined
using the neighboring Tycho-2 stars.  The result is summarized in
table \ref{tab:multicolor}.  The CCD multicolor photometry covered
the fading portion of the light curve.  $B-V$'s were between +0.74 and
+0.81, and $V-I_{\rm c}$'s were between +0.90 and +1.11.  Both relatively
large color indices suggest a significant reddening.  No clear systematic
color variation was observed during the observation period.

\begin{table}
\caption{CCD multicolor photometry.}\label{tab:multicolor}
\begin{center}
\begin{tabular}{ccccc}
\hline\hline
Date (mid UT) &  $B$  &  $V$  & $R_{\rm c}$ & $I_{\rm c}$ \\
\hline
March 16.780 & 12.61 & 11.87 & 11.21 & 10.81 \\
March 17.791 & 12.63 & 11.88 & 11.22 & 10.84 \\
March 24.768 & 13.27 & 12.46 & 11.77 & 11.35 \\
March 31.795 & $\cdots$ & 14.47 & 13.40 & 13.57 \\
April 3.778  & $\cdots$ & 14.53 & $\cdots$ & 13.49 \\
April 8.764  & $\cdots$ & 14.47 & $\cdots$ & 13.54 \\
\hline
\end{tabular}
\end{center}
\end{table}

   Photographic photometry was performed with twin patrol cameras equipped
with a $D$ = 10 cm f/4.0 telephoto lens and unfiltered T-Max 400 emulsions,
located at two sites in Toyohashi, Aichi (KH) and Saku, Nagano (KT).
The passband of observations covers the range of 400--650 nm.  The magnitudes
were derived against comparison stars selected from a GSC quick-$V$ sequence,
whose zero-point adjustment has confirmed using Tycho-2 values.
The overall uncertainty of the calibration and individual
photometric estimates is 0.2--0.3 mag, which will not affect the following
analysis.

\begin{figure*}
  \begin{center}
%    \FigureFile(140mm,95mm){lc.eps}
    \FigureFile(140mm,95mm){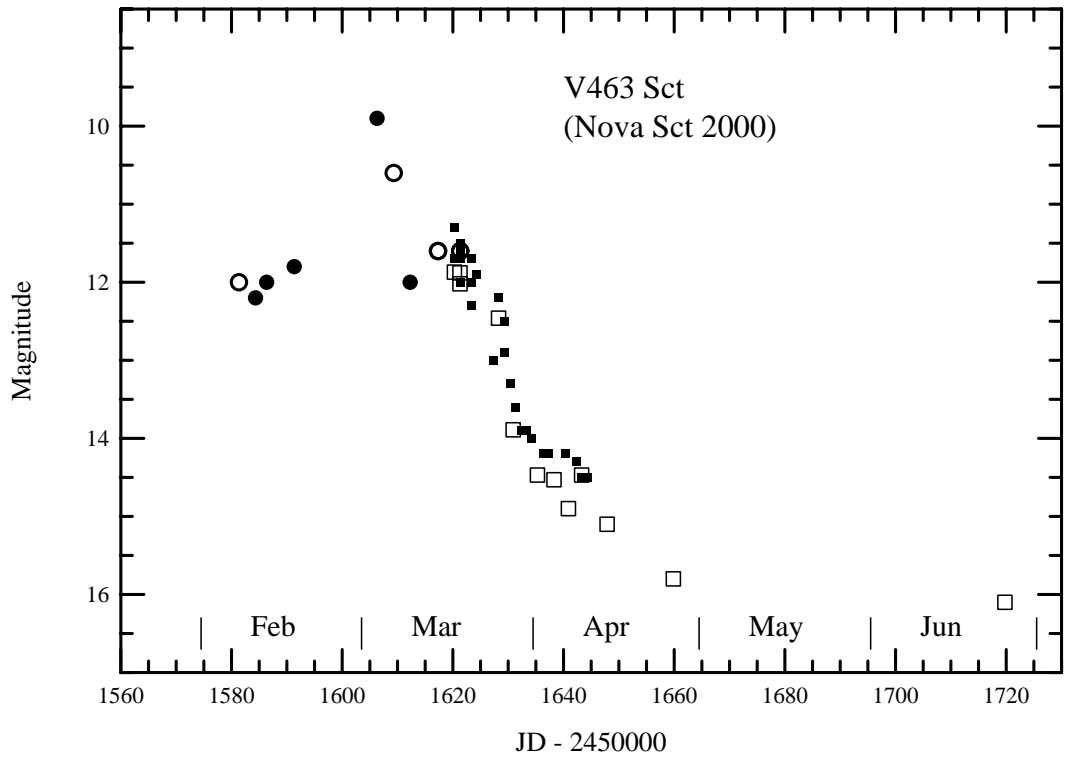}
  \end{center}
  \caption{Light curve of V463 Sct drawn from visual observations
  reported to VSNET (dots), photographic observations (open circles: KH,
  filled circles: KT),
  and CCD $V$-band observations (open squares).}
  \label{fig:lc}
\end{figure*}

   Figure \ref{fig:lc} shows a light curve drawn from visual observations
reported to VSNET, photographic observations, and CCD $V$-band observations.

\section{Spectroscopy}

\begin{figure*}
  \begin{center}
%    \FigureFile(140mm,90mm){spec.eps}
    \FigureFile(140mm,90mm){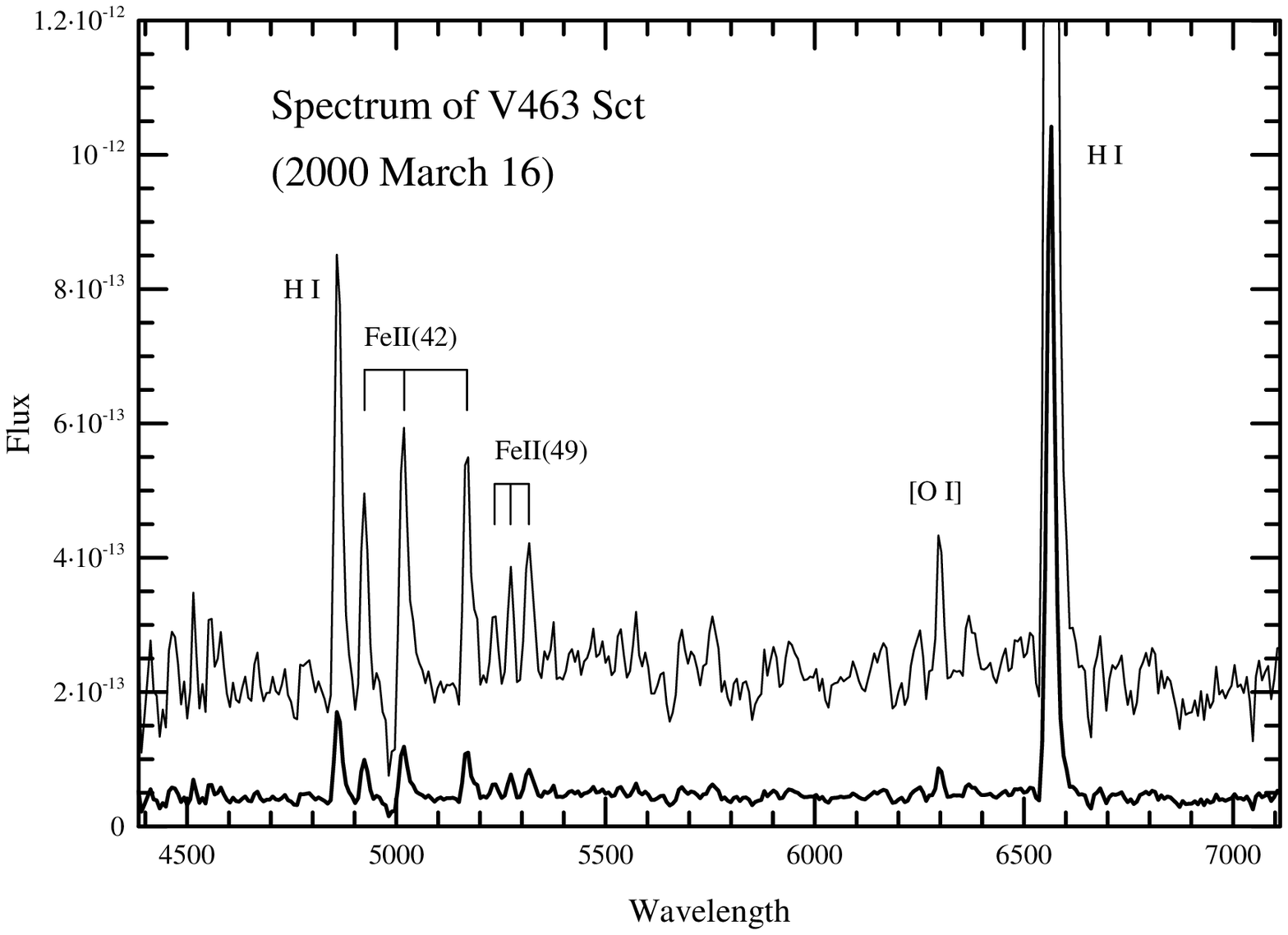}
  \end{center}
  \caption{Spectrum of V463 Sct taken on 2000 March 16.81 UT.
  The unit in flux is erg s$^{-1}$ cm$^{-2}$ \AA$^{-1}$, calibrated using the
  standard star HR 7596.  The flux of the upper (thin line) spectrum has
  been multiplied by 5 in order to show weak lines.
  A strong H$\alpha$ emission line (FWHM = 990 km s$^{-1}$) with a slight
  hint of a P Cyg-type profile is clearly seen.  Fe\textsc{II}
  series emission lines and H$\beta$ emission line are also detected.}
  \label{fig:spec}
\end{figure*}

   MF obtained two spectra (spectral resolution $R$ = 450 at 5500 \AA,
total exposure time 1080 s) on 2000 March 16.81 UT with the FBSPEC-1
spectrograph attached to a 28-cm telescope at Fujii Bisei Observatory.
The data reduction was performed using IRAF package\footnote{
IRAF is distributed by the National Optical Astronomy Observatories.}
and the standard star HR 7596.  Figure \ref{fig:spec} shows an average
of the two spectra.  A strong H$\alpha$ emission line (FWHM = 990 km
s$^{-1}$, refined measurement superseding the value in \cite{uem00v463sct})
with a slight hint of P Cyg-type profile (corresponding a velocity
of 690 km s$^{-1}$) was detected.  Fe\textsc{II}
series emission lines and H$\beta$ emission line were also detected.
The nova is thus thus confirmed to be an Fe\textsc{II} class nova
\citep{wil92novaspec}.

   The line identifications are summarized in Table \ref{tab:lineid}.
The identifications of forbidden lines ([O\textsc{I}] and [N\textsc{II}])
have been confirmed by a comparison with spectroscopic atlases of
novae (\cite{wil91novaspec}; \cite{wil94novaspec}).
The Balmer decrement was steep ($F(H\alpha)/F(H\beta)$ = 7.9).

\begin{table}
\caption{Line identifications.}\label{tab:lineid}
\begin{center}
\begin{tabular}{cccc}
\hline\hline
$\lambda_{\rm obs}$ & E.W.$^*$ & ID & $\lambda$ \\
\hline
4516 & $-$7.5  & Fe\textsc{II} (37) & 4515.3 \\
4555 & $-$16.1 & Fe\textsc{II} (37) & 4555.9 \\
4579 & $-$20.9 & Fe\textsc{II} (37,38,48) & 4582.8, 4576.3, \\
     &         &                  & 4574.8$^\dagger$ \\
4666 & $-$5.3  & Fe\textsc{II} (37) & 4666.8 \\
4861 & $-$76.9 & H$\beta$   & 4861.3 \\
4924 & $-$27.6 & Fe\textsc{II} (42) & 4923.9 \\
5017 & $-$38.5 & Fe\textsc{II} (42) & 5018.9 \\
5169 & $-$50.1 & Fe\textsc{II} (42) & 5169.0 \\
5234 & $-$9.7  & Fe\textsc{II} (49) & 5234.6 \\
5273 & $-$13.5 & Fe\textsc{II} (49) & 5276.0 \\
5317 & $-$20.5 & Fe\textsc{II} (48,49) & 5316.8, 5316.6$^\dagger$ \\
5376 & $-$4.6  & Fe\textsc{II} (49) & 5376.0 \\
5535 & $-$5.6  & Fe\textsc{II} (55) & 5534.8 \\
5572 & $-$5.7  & [O\textsc{I}]      & 5577$^\ddagger$ \\
5681 & $-$10.4 & N\textsc{II}       & 5679$^\ddagger$ \\
5756 & $-$21.4 & [N\textsc{II}]     & 5755 \\
5897 & $-$3.5, $-$7.7$^\S$ & Na\textsc{I} & 5890 \\
6246 & $-$14.9 & Fe\textsc{II} (74) & 6240.1, 6247.6$^\dagger$ \\
6299 & $-$16.9 & [O\textsc{I}]      & 6300 \\
6365 & $-$4.2  & [O\textsc{I}]      & 6364 \\
6454 & $-$1.8  & Fe\textsc{II} (74) & 6456.4 \\
6564 & $-$427.8 & H$\alpha$  & 6562.8 \\
\hline
 \multicolumn{4}{l}{$^*$ Equivalent width in \AA.} \\
 \multicolumn{4}{l}{$^\dagger$ Blended.} \\
 \multicolumn{4}{l}{$^\ddagger$ Uncertain identification.} \\
 \multicolumn{4}{l}{$^\S$ Decomposed doublet components.} \\
\end{tabular}
\end{center}
\end{table}

\section{Discussion}

\subsection{Light Curve and Absolute Magnitude}\label{sec:absmag}

   At a first look, the rapid evolution of the light curve around in
early March seems to suggest that the brightest observation by KT on
March 2 was obtained close to the epoch of the maximum.  Following this
interpretation, we obtain $t_2$ = 6 d based on KT's subsequent
observations using the same system.  This value would indicate that V463 Sct
belongs to very fast novae \citep{due87novaatlas}.
By applying a recent calibration between the rate of decline and absolute
magnitude at maximum \citep{dellaval95novaabsmag}, we obtain the maximum
$M_{\rm V}$ = $-$8.9 $\pm$ 0.5.  Since photographic observations and
visual observations have been confirmed to agree within $\sim$0.3 mag,
we can then safely adopt apparent maximum magnitude of $V$ = 9.9.
By adopting a mean $B-V$ of +0.75 during the early decline phase
and the intrinsic color $B-V$ = $-0.02 \pm 0.04$ for novae 2 mag
below the maximum \citep{vandenber87novaUBV}, we obtained
$E_{B-V}$ = +0.8 $\pm$ 0.1.  The observed constancy of the $B-V$'s
during the observation precludes a possible effect of a reddening pulse
which occurs in some novae around their optical maxima
\citep{vandenber87novaUBV}.
Using the generally adopted relation $A_{\rm V} = 3.1 E_{B-V}$ and
$E_{B-V}$, we can derive a distance modulus of 16.3 $\pm$ 0.6.
This value is, however, too large to accept.

   The other interpretation is that the ``maximum" caught on March 2
is better understood as a delayed flare-like maximum sometimes seen
in very slow novae [e.g.
HR Del (\cite{ter70hrdellvvul}, \cite{ter74hrdel}, \cite{dre77hrdel}),
V723 Cas (\cite{ohs96v723cas}, \cite{mun96v723cas}, \cite{iij98v723cas}),
V1548 Aql (\cite{kat01v1548aql}),
and DO Aql (\cite{vog28doaql}, \cite{bey29doaql})
].  The measured line width (FWHM = 990 km s$^{-1}$) seems to
support this analogy [the FWHMs of V1548 Aql and V723 Cas were reported
to be 1100 and 600 km s$^{-1}$, respectively
(\cite{she00v1548aqliauc}, \cite{dellaval95v723casiauc})]\footnote{
  A care, however, should be taken to make a direct comparison of line
  widths derived at different stages of outbursts.  One should be also
  careful in interpreting the velocity in V463 Sct because a higher velocity
  component could have been undetected in the present low-resolution
  spectra.
}
The flat portion of the light curve around $V$ = 12.0, following the first
positive detection $\sim$ 24 d before March 2, is reasonably interpreted
as a premaximum halt.  The exact duration of the premaximum halt was not
determined because of the lack of observations (around the solar conjunction)
between 1999 November 10 and 2000 February 6.  On 1999 November 10,
the object was fainter than $V$ = 14.0.
Following this interpretation, the linear portion of the later evolution
of the light curve would better represent the characteristic decline
rate.  The linear decline rate since March 2 has been estimated to
correspond to $t_2$ = 15 $\pm$ 3 d.  If we adopt $t_2$ = 15 and the
relation in \citet{dellaval95novaabsmag}, $M_{\rm V}$ becomes only 0.5 mag
fainter than the case of $t_2$ = 6 d, which yields a still unacceptably
large distance modulus of 15.8 $\pm$ 0.6.  These estimates suggest
that the speed of the linear decline part is not a useful indicator of
the absolute magnitude of this nova.

   The lack of large color changes during the fading stage seems to
preclude a possibility of a dust-forming episode (see \citet{due81novatype}
for representative color changes).

   The nova is located on the edge of Lynds 387.  The large interstellar
absorption suggests that the nova is behind this molecular cloud.

\subsection{Quiescent Counterpart}

   Since there is a considerable uncertainty in estimating the decline
rate, we have adopted $M_{\rm V}$ = $-$7.7 (typical maximum absolute
magnitude for classical novae) as in \citep{lil87novarate}.
Even if we adopt an extreme value of $M_{\rm V}$ = $-$8.9, the following
discussion will not be considerably affected.
The absence of a quiescent optical counterpart down to $V \sim$ 20
corresponds to an upper limit of $M_{\rm V}$ = 2 ($M_{\rm V}$ = 1 for
the $M_{\rm V}$(max) = $-$8.9 case).  This upper limit
is reasonable for a CV-type quiescent counterpart of a classical nova
(e.g. \citet{war86NLabsmag} showed that most of post-outburst classical
novae show $M_{\rm V}$ fainter than +2).
This limit and the non-detection on 2MASS images,
however, precludes the possibility of a red giant secondary, as in
symbiotic (recurrent) novae \citep{anu99RN}.  The present upper limit
is insufficient to exclude the possibility of a CV-type recurrent nova
\citep{hac01RN}.

\subsection{Premaximum Halt}\label{sec:premaximum}

   As shown in subsection \ref{sec:absmag}, V463 Sct is clearly
exhibited the fastest evolution among the recorded novae with a late-time
flare-like maximum and a flat premaximum
(premaximum halt), even considering the unavoidable uncertainty
in determining $t_2$.  An literature search in \citet{GalacticNovae} and
\citet{due81novatype}, well-observed fast to moderately fast novae from
\citet{due87novaatlas}, and observations of recent novae in VSNET, VSOLJ
and AFOEV observations has not revealed a similar example.  This lack of
a similar example may have been a result of a selection effect that
novae are usually discovered near their optical maxima and very few
objects were recorded in their premaximum stage.

   Although the origin of the phenomenon is not
yet well understood, it is well known that the duration of a premaximum
halt depends on the nova speed class: in slow novae, the typical duration
is a few to several days; in fast novae, the typical duration is a few
hours (\cite{ClassicalNovae}, p.4).
In some slow novae (HR Del, V1548 Aql, V723 Cas, DO Aql, see
subsection \ref{sec:absmag} for the references), the durations can be
up to several months.  All these exceptional slow novae are known to
belong to the ``slowest" novae (there is even an argument that V723 Cas
may be a symbiotic nova, cf. \cite{mun96v723cas}).  The occurrence of
at least 24 d-long premaximum halt in a rapidly evolving nova
(V463 Sct) is quite exceptional.  This observation suggests that the
occurrence of a long
premaximum halt and a late-time flare-like maximum is not a unique
character of the ``slowest" novae, but a more general phenomenon spreading
over a wider range of nova speed classes than has been previously
believed.

\begin{figure}
  \begin{center}
%    \FigureFile(88mm,120mm){comp.eps}
    \FigureFile(88mm,120mm){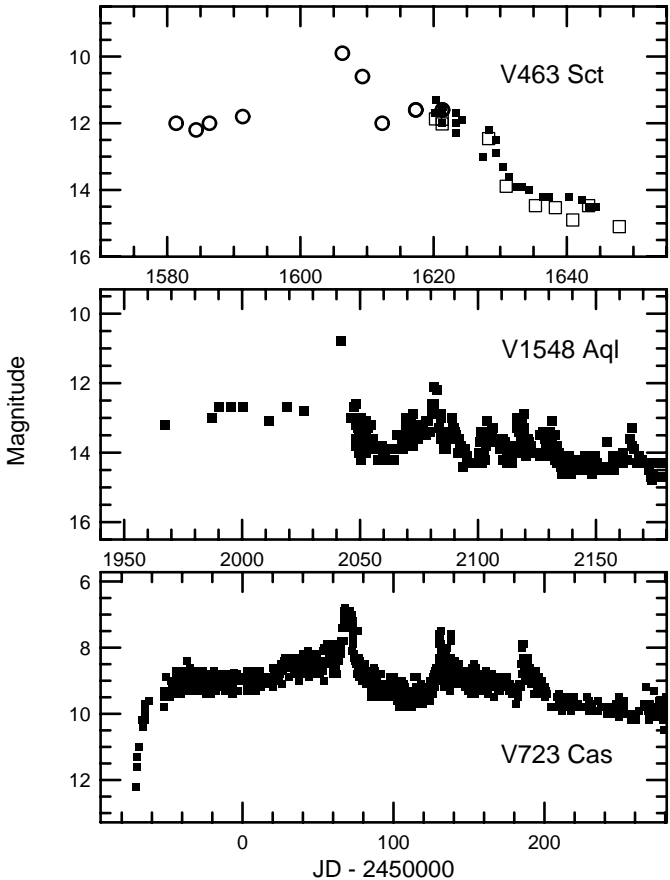}
  \end{center}
  \caption{Comparison of three recent novae with a prominent premaximum
  halt.  The data for V463 Sct are the same as in figure \ref{fig:lc}.
  The data for V1548 Aql and V723 Cas are from reports to VSNET.
  The scales of the axes are different between the objects.  The quick
  evolution of V463 Sct is evident.
  }
  \label{fig:comp}
\end{figure}

   Figure \ref{fig:comp} shows a comparison of three recent
novae with a prominent premaximum halt.  The data for V463 Sct are the
same as in figure \ref{fig:lc}.  The data for V1548 Aql and V723 Cas are
from reports to VSNET.  The scales of the axes are different between the
objects.  The quick evolution of V463 Sct is evident.

   \citet{fri92hrdel} studied the past spectroscopic observations of
HR Del during its long premaximum stage.  \citet{fri92hrdel} showed the
presence of an almost stationary photosphere with very low velocities,
unlike for the majority of classical novae.  \citet{fri92hrdel} suggested
that the conditions of thermonuclear runaway was only marginally satisfied
in HR Del.  This interpretation is compatible with the measured white
dwarf mass in HR Del (0.52 $M_{\odot}$: \cite{bru82hrdel}; 0.595 $M_{\odot}$:
\cite{kur88hrdel}) combined with models of thermonuclear runaways
\citep{pri95novaevolution}.
This interpretation seems to apply to similar very slow novae with
a long premaximum halt and delayed maximum.  \citet{ori93localTNR}
proposed that a premaximum halt may be explained as an effect of
local thermonuclear runaways, but this possibility has not been
examined in more detail.

   If the interpretation of \citet{fri92hrdel} also applies to V463 Sct,
the rapid subsequent evolution should meet a difficulty because rapidly
evolving novae would require a more massive white dwarf
(see e.g. \cite{pri95novaevolution}).
If this marginal condition is met in V463 Sct, this object should have
a high mass-transfer rate.  The uniqueness of V463 Sct in that it showed
both a slow-nova type long premaximum halt and rapid photometric evolution
is expected to provide a most stringent test for this interpretation.
Future quiescent observation of V463 Sct to determine the white dwarf
mass and an accretion rate would therefore provide an unique opportunity
in testing the interpretation that premaximum halts reflect a marginally
satisfied TNR condition.

\subsection{Spectroscopy}

   The ``textbook" development of the optical spectra of novae has been
described by various authors (e.g. \cite{GalacticNovae};
\cite{ClassicalNovae}, Chap. 1.4; \cite{wil92novaspec}; \cite{wil91novaspec};
\cite{wil94novaspec}).
The post-maximum evolution of nova spectra is known to be complex
(\cite{wil92novaspec}; \cite{wil94novaspec}).  \citet{wil92novaspec}
described that novae at early post-outburst decline can be divided into
two classes: Fe\textsc{II} novae and He/N novae based on the predominant
lines.  \citet{wil92novaspec} reported that Fe\textsc{II} novae, to which
V463 Sct belongs, tend to have narrower lines (i.e. smaller expansion
velocity) and relatively early appearance of [O\textsc{I}] lines.
The [O\textsc{I}] ($\lambda\lambda$ 6300, 6364 \AA) lines, however, usually
appear later than the evolution of Fe\textsc{II} lines.  For example,
\citet{wil94novaspec} showed that the [O\textsc{I}] ($\lambda\lambda$ 6300,
6364 \AA) lines were not strong 14 d after maximum in Nova LMC 1991,
11 d after maximum of V351 Pup = Nova Pup 1991, 11 d after maximum of
V2264 Oph = Nova Oph 1992 No. 1.  All of these novae are fast novae with
$t_3$ between 8 and 26 d.

   A spectrum of V463 Sct is shown in figure \ref{fig:spec}.  According
to \citet{wil94novaspec}, it usually takes 20--30 d in fast novae before
the [O\textsc{I}] lines become comparable in strength to V463 Sct
at this epoch.  The early (only 14 d after the March 2 maximum) appearance
of these lines in V463 Sct is thus rather exceptional among Fe\textsc{II}
class novae. The spectrum is also similar to a spectrum of the moderately
fast novae V1425 Aql (Nova Aql 1995) taken 50.5 d after the maximum
\citep{kam97v1425aql}, V443 Sct (Nova Sct 1989) 72 d after
the maximum \citep{anu92v443sct} and V1819 Cyg (Nova Cyg 1986) 92 d after
the maximum \citep{whi89v1819cyg}.  These comparisons suggest that V463 Sct
showed a moderately evolved spectrum only 14 d after the March 2 maximum.

   According to \citet{wil92novaspec}, the Fe\textsc{II} lines are considered
to be formed in high-density winds.  Since neutral oxygen is not expected
in a homogeneous, high-temperature nebula in a nova, the presence of
the [O\textsc{I}] lines is usually considered to be a signature of
density inhomogeneities which are sufficiently large to prevent complete
ionization \citep{wil92novaspec}.  The presence of the Fe\textsc{II} and
[O\textsc{I}] lines suggests that substantial amount of material had been
ejected.  Since the appearance of the [O\textsc{I}] lines in V463 Sct seems
to be earlier than in other Fe\textsc{II} class novae, the mass ejection
responsible for the [O\textsc{I}] lines may have taken place earlier
than the March 2 maximum.  Since the elapsed time (38 d) since the earliest
positive detection (February 6) is comparable to the epochs of appearance
of the [O\textsc{I}] lines in fast to moderately fast Fe\textsc{II} class
novae, the presence of a long premaximum halt may be responsible for
this ejection.

   The steep Balmer decrement is consistent with a relatively large
reddening (subsection \ref{sec:absmag}).

\subsection{Importance of Rapid Identification}

   As described in section \ref{sec:discovery}, the nova was original
considered as a red variable based on a possible identification with
an IRAS source.  Such a technique using catalog correlations to screen
out large-amplitude red variables among nova candidates has been widely
used.  The present example clearly demonstrate a caveat of this technique
in a very crowded region (V463 Sct indeed lied within an error ellipse
of the IRAS source).  It is thus highly advisable to check all materials
(DSS, 2MASS, past archival photographs) before completely ruling out
the possibility of a nova even if the suspect object is close to
an IRAS source.  Rapid accurate astrometry is confirmed to be very
helpful in accurately identifying the counterpart on DSS images, leading
to a correct interpretation of a new object.  Finally, a rapid circulation
of discovery information through a world-wide network such as VSNET
is proven to be very effective even when the nature of the new object
is not clarified.  Without such a medium, a chance in making early epoch
observations could have been easily lost (for an unfortunate example, see
\cite{kat01v1178sco}) especially in fast novae.

\vskip 3mm

We are grateful to many VSNET, VSOLJ and AFOEV observers who have reported
vital observations.
This work is partly supported by a grant-in aid [13640239 (TK),
14740131 (HY)] from the Japanese Ministry of Education, Culture, Sports,
Science and Technology.
Part of this work is supported by a Research Fellowship of the
Japan Society for the Promotion of Science for Young Scientists (MU).
This research has made use of the Digitized Sky Survey producted by STScI, 
the ESO Skycat tool, and the VizieR catalogue access tool.


\begin{thebibliography}{}

\bibitem[Anupama et~al.(1992)]{anu92v443sct}
  Anupama, G.~C., Duerbeck, H.~W., Prabhu, T.~P., \& Jain, S.~K.\ 1992, \aap,
  263, 87

\bibitem[Anupama, Mikolajewska(1999)]{anu99RN}
  Anupama, G.~C., \& Mikolajewska, J.\ 1999, \aap, 344, 177

\bibitem[Beyer(1929)]{bey29doaql}
  Beyer, M.\ 1929, Astron. Nachr., 235, 427

\bibitem[Bode, Evans(1989)]{ClassicalNovae}
  Bode, M.~F., \& Evans, A.\ 1989, Classical Novae (Alden Press: Oxford)

\bibitem[Bruch(1982)]{bru82hrdel}
  Bruch, A.\ 1982, \pasp, 94, 916

\bibitem[della Valle, Livio(1995)]{dellaval95novaabsmag}
  della Valle, M., \& Livio, M.\ 1995, \apj, 452, 704

\bibitem[della Valle et~al.(1995)]{dellaval95v723casiauc}
  della Valle, M., Marchiotto, W., \& Lercher, G.\ 1995, \iaucirc, 6214

\bibitem[Drechsel et~al.(1977)]{dre77hrdel}
  Drechsel, H., Rahe, J., Duerbeck, H.~W., Kohoutek, H.~W., \& Seitter, W.~C.\
  1977, \aaps, 30, 323

\bibitem[Duerbeck(1981)]{due81novatype}
  Duerbeck, H.~W.\ 1981, \pasp, 93, 165

\bibitem[Duerbeck(1987)]{due87novaatlas}
  Duerbeck, H.~W.\ 1987, \ssr, 45, 1

\bibitem[Friedjung(1992)]{fri92hrdel}
  Friedjung, M.\ 1992, \aap, 262, 487

\bibitem[Hachisu, Kato(2001)]{hac01RN}
  Hachisu, I., \& Kato, M.\ 2001, \apj, 558, 323

\bibitem[Haseda(2000)]{has00v463sct}
  Haseda, K.\ 2000, \iaucirc, 7382

\bibitem[Iijima et~al.(1998)]{iij98v723cas}
  Iijima, T., Rosino, L., \& della Valle, M.\ 1998, \aap, 338, 1006

\bibitem[Kamath et~al.(1997)]{kam97v1425aql}
  Kamath, U.~S., Anupama, G.~C., Ashok, N.~M., \& Chandrasekhar, T.\ 1997, \aj,
  114, 2671

\bibitem[Kato, Fujii(2001)]{kat01v1178sco}
  Kato, T., \& Fujii, M.\ 2001, Inf. Bull. Variable Stars, 5150

\bibitem[Kato et~al.(2001)]{kat01hvvir}
  Kato, T., Sekine, Y., \& Hirata, R.\ 2001, \pasj, 53, 1191

\bibitem[Kato, Takamizawa(2001)]{kat01v1548aql}
  Kato, T., \& Takamizawa, K.\ 2001, Inf. Bull. Variable Stars, 5100

\bibitem[K"{u}rster, Barwig(1988)]{kur88hrdel}
  K"{u}rster, M., \& Barwig, H.\ 1988, \aap, 199, 201

\bibitem[Liller, Mayer(1987)]{lil87novarate}
  Liller, W., \& Mayer, B.\ 1987, \pasp, 99, 606

\bibitem[Munari et~al.(1996)]{mun96v723cas}
  Munari, U., Goranskij, V.~P., Popova, A.~A., Shugarov, S.~Y., Tatarnikov,
  A.~M., Yudin, B.~F., Karitskaya, E.~A., Kusakin, A.~V., {et~al.}\ 1996, \aap,
  315, 166

\bibitem[Ohsima et~al.(1996)]{ohs96v723cas}
  Ohsima, O., Akazawa, H., \& Ohkura, N.\ 1996, Inf. Bull. Variable Stars,
  4295

\bibitem[Orio, Shaviv(1993)]{ori93localTNR}
  Orio, M., \& Shaviv, G.\ 1993, \apss, 202, 273

\bibitem[Payne-Gaposchkin(1957)]{GalacticNovae}
  Payne-Gaposchkin, C.\ 1957, The Galactic Novae (North-Holland: Amsterdam)

\bibitem[Prialnik, Kovetz(1995)]{pri95novaevolution}
  Prialnik, D., \& Kovetz, A.\ 1995, \apj, 445, 789

\bibitem[Shemmer(2001)]{she00v1548aqliauc}
  Shemmer, O.\ 2001, \iaucirc, 7628

\bibitem[Starrfield(1999)]{sta99novareview}
  Starrfield, S.\ 1999, Phys. Rep., 311, 371

\bibitem[Starrfield, Sparks(1987)]{sta87novareview}
  Starrfield, S., \& Sparks, W.~M.\ 1987, \apss, 131, 379

\bibitem[Starrfield et~al.(2000)]{sta00novareview}
  Starrfield, S., Truran, J.~W., \& Sparks, W.~M.\ 2000, New Astron. Rev.,
  44, 81

\bibitem[Terzan(1970)]{ter70hrdellvvul}
  Terzan, A.\ 1970, \aap, 5, 167

\bibitem[Terzan et~al.(1974)]{ter74hrdel}
  Terzan, A., Bally, M., \& Durand, A.\ 1974, \aaps, 15, 107

\bibitem[Uemura et~al.(2000)]{uem00v463sct}
  Uemura, M., Kato, T., Yamaoka, H., Fujii, M., Quinn, J., Garnavich, P., \&
  Takamizawa, K.\ 2000, \iaucirc, 7382

\bibitem[van~den Bergh, Younger(1987)]{vandenber87novaUBV}
  van~den Bergh, S., \& Younger, P.~F.\ 1987, \aaps, 70, 125

\bibitem[Vogt(1928)]{vog28doaql}
  Vogt, H.\ 1928, Astron. Nachr., 232, 269

\bibitem[Warner(1986)]{war86NLabsmag}
  Warner, B.\ 1986, \mnras, 222, 11

\bibitem[Warner(1995)]{war95book}
  Warner, B.\ 1995, Cataclysmic Variable Stars (Cambridge: Cambridge
  University Press)

\bibitem[Whitney, Clayton(1989)]{whi89v1819cyg}
  Whitney, B.~A., \& Clayton, G.~C.\ 1989, \aj, 98, 297

\bibitem[Williams(1992)]{wil92novaspec}
  Williams, R.~E.\ 1992, \aj, 104, 725

\bibitem[Williams et~al.(1991)]{wil91novaspec}
  Williams, R.~E., Hamuy, M., Phillips, M.~M., Heathcote, S.~R., Wells, L., \&
  Navarrete, M.\ 1991, \apj, 376, 721

\bibitem[Williams et~al.(1994)]{wil94novaspec}
  Williams, R.~E., Phillips, M.~M., \& Hamuy, M.\ 1994, \apjs, 90, 297

\bibitem[Zhao et~al.(1994)]{zha94lqsgr}
  Zhao, P., McClintock, J.~E., \& Landolt, A.~U.\ 1994, \iaucirc, 6018

\end{thebibliography}
\end{document}